\documentclass{emulateapj}
\usepackage{amssymb,amsmath,bm}
\begin{document}

\title{Imprint of Inhomogeneous Reionization on the Power Spectrum \\
of Galaxy Surveys at High Redshifts}

\author{Daniel Babich\altaffilmark{1,2} and Abraham Loeb\altaffilmark{2,3}}

\altaffiltext{1}{Department of Physics, Harvard University, Cambridge, MA
02138; babich@physics.harvard.edu.}  

\altaffiltext{2}{Harvard-Smithsonian Center for Astrophysics, 60 Garden
Street, Cambridge, MA 02138} 

\altaffiltext{3}{Department of Astronomy, Harvard University, Cambridge, MA
02138; aloeb@cfa.harvard.edu.}

\begin{abstract}
We consider the effects of inhomogeneous reionization on the distribution
of galaxies at high redshifts.  Modulation of the formation process of the
ionizing sources by large scale density modes makes reionization inhomogeneous 
and introduces a spread to the reionization times of different regions with the 
same size. After sources photo-ionize and heat these regions to a temperature 
$\ga 10^4$K at different times, their temperatures evolve as the ionized 
intergalactic medium (IGM) expands.  The varying IGM temperature makes the
minimum mass of galaxies spatially non-uniform with a fluctuation amplitude
that increases towards small scales. These scale-dependent fluctuations
modify the shape of the power spectrum of low-mass galaxies at high
redshifts in a way that depends on the history of reionization. The
resulting distortion of the primordial power spectrum is significantly
larger than changes associated with uncertainties in the inflationary
parameters, such as the spectral index of the scalar power spectrum or the
running of the spectral index. Future surveys of high-redshift galaxies
will offer a new probe of the thermal history of the IGM but might have a 
more limited scope in constraining inflation.

\end{abstract}


\section{Introduction}

The reionization of the cosmic neutral hydrogen, left over from the big
bang, was an intrinsically inhomogeneous process \citep{Barkana04,
Furlanetto05}. The basic cause of this inhomogeneity is modulation by large
scale structure of the formation process of the ionizing sources. The
scatter in the local matter density produces corresponding scatter in
physical observables during the epoch of reionization.  In particular,
modulation of the formation of the first sources by large scale density
modes will introduce scatter into the redshift at which reionization is
completed in different regions due to the scatter in local matter density
\citep{Wyithe04}. This implies that the redshift of reionization is not a
single number that is universally applicable to all regions of the universe,
but rather follows a distribution which strongly depends on the physical
size of the region under consideration \citep{Barkana04,Wyithe05a}. By a
fixed redshift, different regions of a given size expand and cool by
different amounts since they were reheated by reionization at different
times.  Inhomogeneous reionization leaves scars on the IGM in the form of a
scale-dependent distribution of temperatures and densities.

The fluctuations in the IGM temperature affect galaxy formation by
increasing the minimum mass of a dark matter halo that would accrete gas
from the photo-heated IGM \citep{Haiman96,Dijkstra04}. Since the
temperature fluctuations are scale dependent, the variance of the
distribution of galaxies will also be strongly scale dependent. This will
strongly influence the power spectrum of fluctuations in the galaxy
distribution and alter its shape relative to the underlying power spectrum
of the dark matter. Previous treatments of the reheating feedback
\citep{Barkana00, Efstathiou92} assumed that the photo-heating of the IGM
and the corresponding suppression of galaxy formation was uniform.  In this
paper we explore the implications of the non-uniformity of this process.

Future observations of the high redshift galaxy power spectrum, $P(k)$, may
be used to constrain cosmological parameters. Since the scale at which the
dark matter power spectrum becomes non-linear is smaller at high redshift,
the high-redshift galaxy surveys would be able to probe smaller scales
(i.e. modes with a higher wavenumber $k$) than their low-redshift
counterparts such as {\it SDSS} or {\it 2dF}. The combination of the small
scale data along with low redshift galaxy surveys and cosmic microwave
background (CMB) measurements, may significantly improve our ability to
constrain inflationary parameters. This is particularly true for the
spectral index $n \equiv d\log{P(k)}/d \log{k}$, and the running of the
spectral index $\alpha \equiv d n /d \log{k} $, since the small scale
observations increases the range of $k$--modes over which these inflationary
parameters can be constrained.

The {\it Cosmic Inflation Probe} ({\it CIP}) is a proposed mission to
undertake such a high redshift galaxy redshift survey
\citep{Melnick05}. {\it CIP} is designed to be a space-based large area
galaxy redshift survey between $2.5~\mu$m and $5~\mu$m, that would observe
$\sim 10^7$ galaxies in H$\alpha$ emission within the redshift interval $z
\sim 3$ to $6.5$. In such a survey, the galaxy power spectrum could be
determined to an exquisite precision of $\la 1\%$ over the wavenumber
interval of $k = 0.01$ to $1$ comoving $\mbox{ Mpc}^{-1}$.  Assuming linear
bias these observations could potentially constrain the spectral index to
better than $\Delta n = \pm 0.002$. However, we will show in this paper
that the corrections to the galaxy power spectrum imprinted by the scale
dependent fluctuations of the IGM temperature after reionization are far
greater than the existing uncertainties in the inflationary parameters.

The organization of the paper is as follows. In \S 2 we discuss the manner
by which large scale modes would introduce scatter into the redshift of
reionization, and how this will impact galaxy formation. In \S 3 we present
numerical results for the imprint of fluctuations in the IGM temperature on
the galaxy power spectrum. Finally, \S 4 summarizes our main conclusions.
We adopt the standard $\Lambda$CDM cosmological model consistent with the
{\it WMAP} data \citep{Spergel03}, with the parameters $\Omega_b = 0.044$,
$\Omega_m = 0.27$, $\Omega_v = 0.73$, $n = 1$, $\sigma_8 = 0.9$ and $h =
0.72$. All distances and wavenumbers throughout the paper are in comoving
coordinates.

\section{Inhomogeneous Reionization}

As discussed in \S 1, the process of reionization was inhomogeneous
due to the modulation of structure formation by large
scale density modes. The associated fluctuations in the redshift of
reionization produced corresponding fluctuations in the temperature and
density of the photo-ionized IGM. Since different regions were reionized at
different times, they cool by different amounts and therefore
obtain a distribution of temperatures at any redshift slice of the
universe. We will assume that all reionizing sources had the same spectrum
and therefore photo-heated the local IGM to the same temperature. It is the
distribution of reionization redshifts, not the scatter in the photo-heated
temperatures, that is assumed to produce the distribution in IGM
temperatures in our formulation\footnote{Other sources of temperature
fluctuations, such as spatial variations in the spectrum of the ionizing
sources (resulting from variations in the relative contributions from
quasars or stars, for example) would only strengthen our basic conclusion
that the imprint of inhomogeneous reionization on $P(k)$ is far greater
than the uncertainty associated with the inflationary initial conditions.}
This temperature distribution produces scatter in physical observables in
the IGM, such as the galaxy formation rate.

\subsection{Fluctuations in the Redshift of Reionizaton}

Any region of size $R$ in the universe has a probability distribution for
its local matter density which will change the redshift at which the
collapsed fraction of matter in the region exceeds some threshold and makes
the region reionized \citep{Barkana04}. This threshold is determined by the
astrophysics of the problem, such as the star formation efficiency, the
radiation spectrum, the escape fraction of ionizing photons from galaxies,
as well as the clumping factor of the high redshift IGM
\citep{Furlanetto04a}. However, given the redshift of
reionization\footnote{This redshift corresponds to when an average
($\delta_R = 0$) region will be ionized. We assume that voids ($\delta_R <
0$) are ionized by nearby overdense regions at $z_R$ (so this is not the
average but rather the completion redshift of reionization).}, $z_R$,
\citet{Barkana04} showed that fluctuations around this redshift are
independent of these astrophysical complications.  The large-scale
polarization anisotropies of the CMB \citep{Kogut03} and the scale of the
ionized regions around quasars at $z\ga 6$
\citep{Wyithe04,Mesinger04,Wyithe05b} indicate that the redshift of
reionization is most likely between $z_R \sim 6$ and $15$.  Long wavelength
density fluctuations contribute to the background density and alter the
rate of evolution within a region with respect to the background. This
implies that the time at which the collapsed fraction of the perturbed
region reaches a reionization threshold will be randomly shifted with
respect to the time at which an unperturbed region reaches the threshold
\citep{Bardeen86,Barkana04}.  The collapsed fraction in the modulated
region can be equated to the collapsed fraction in an unperturbed region by
shifting the redshift of the perturbed region by $\delta z$,
\begin{equation}\label{collapsed}
   F\left(\frac{\delta_c(z_R+\delta z) - \delta_R}{\sqrt{2(\sigma^2(R_{\rm min})
   - \sigma^2(R))}}\right) = F\left( \frac{\delta_c(z_R)}{\sqrt{2\sigma^2(R_{\rm min})}}\right),
\end{equation}
where $\delta_R$ is the overdensity of the region under consideration of
radius $R$ and $\delta_c(z)$ is the collapse threshold at redshift $z$.
The mass variance, namely the root-mean-square fluctuation in mass of
regions of size $R$, is defined as
\begin{equation}\label{mass_variance}
   \sigma^2(R) = \int \frac{k^2 dk}{2\pi^2} P(k) \tilde{W}^2(kR),
\end{equation}
where $\tilde{W}(x)$ is the Fourier Transform of the window function. We
adopt a top-hat window function in real space of radius $R$ 
with the Fourier Transform
\begin{equation}
   \tilde{W}(kR) = \frac{3j_1(kR)}{kR}.
\end{equation}

There is a minimum mass of dark matter halos into which the cosmic gas can
collapse, dissipate, fragment and eventually form stars \citep{White78,
Rees77}. In fact there are two independent minimum masses: {\it (i)} the
Jeans mass, which delineates the mass scale when gravitational forces
exceed baryonic pressure support; {\it (ii)} the cooling mass, below which
the gas cooling time is longer than the halo dynamical time and
fragmentation into stars is not possible. Depending on the thermal state of
the IGM, either mass scale will determine the minimum galaxy mass. In the
pre-reionization universe, where the IGM is extremely cold and the minimum
mass is set by the requirement that the cooling time be less than the
dynamic time. While the original reionizing sources can be formed through
dissipation and fragmentation caused by H$_2$ cooling, feedback effects may
have photodissociated much of the H$_2$ in the universe and required future
dissipation and fragmentation to proceed by atomic hydrogen line cooling
\citep{Haiman97}. The minimum virial temperature for atomic Hydrogen
cooling is $\sim 10^4$ K, this consideration sets the minimum virial radius,
$R_{\rm min}$, in equation (\ref{collapsed}). 

Once the IGM is reionized and photo-heated to a characteristic temperature
of $\sim 2\times 10^4$ K, the Jeans mass will exceed the cooling mass. In
fact, we will see in the next section that the Jeans mass is replaced by
the filtering mass, which incorporates the entire IGM thermal
history. Nevertheless, we will assume that the initial minimum scale for
ionizing sources is given by the requirement that the halo can through
atomic Hydrogen line transitions.

Equating the arguments in equation (\ref{collapsed}), we can solve for the
fluctuation in the reionization redshift \citep{Barkana04}
\begin{equation}\label{delta_z}
  \delta z=\frac{\delta_R}{\delta_0}-(1+z_R)\left[1-(1-\frac{\sigma^2(R)}
{\sigma^2(R_{\rm min})})^{1/2}\right],
\end{equation}
where $\delta_0 = \delta_c(z)/(1+z)$ is approximately constant at high-redshift.
Since $\delta_R$ is a random variable drawn out of a Gaussian probability
distribution with zero mean and variance $\sigma^2(R)$, $\delta z$ is also
a random variable. The relation between $\delta_R$ and $\delta z$ is a
linear transformation, and so the distribution of $\delta z$ is
Gaussian, with a different variance. In this paper we will use Monte-Carlo
sampling to calculate the distribution of physical quantities, such as
reionization redshift, gas temperature and fluctuations in the galaxy
number density. Throughout the paper we ensure that our mass density
fluctuations are always physical ($\delta_R \ge -1$) and uncollapsed
($\delta_R < \delta_c(z_G)$) by discarding regions that violate either of
these conditions. We also assume that voids, regions with $\delta_R < 0$,
will be photo-ionized by neighboring overdense regions; and so all voids
will be reionized at $z_R$.

\begin{figure}
\plotone{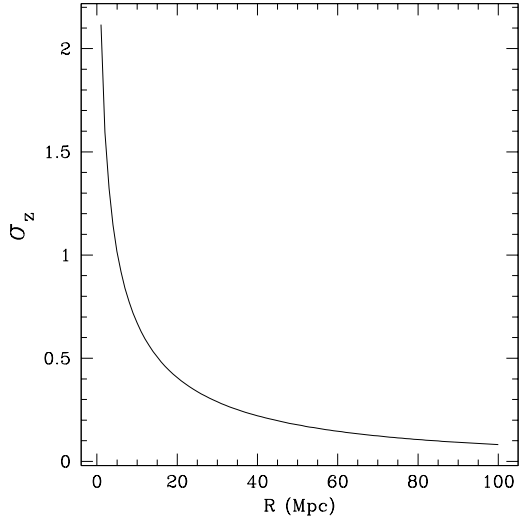}
\caption{The standard deviation of the fluctuations in the redshift of
reionization, $\sigma_z$, as a function of region size.}
\label{dz}	
\end{figure}

Figure \ref{dz} shows the standard deviation of the fluctuations in the
redshift of reionization as a function of region size. As the size of a
region increases the scatter in reionization redshift fluctuations
decreases. On scale $R < 20$ Mpc the fluctuations in the redshift of 
reionization become substantial.

\subsection{High Redshift Galaxy Formation}

The baryons traced the temperature of the CMB with $T \propto (1+z)$ until
they thermally decoupled from it at $z \sim 140$ \citep{Loeb04}, after
which they cooled adiabatically with $T\propto (1+z)^{2}$ until the first
galaxies formed. These galaxies produced photons above the Lyman limit
($13.6$ eV) that ionized and photo-heated the IGM to a temperature set by
the slope of their ionizing spectrum. If stars were the dominant source of
reionization, then the IGM was heated to a temperature $\sim (1$--$2)
\times 10^4$ K (with the precise value depending on the initial mass
function of stars). After being reionized the gas cooled predominantly due
to adiabatic cooling through its Hubble expansion, but at high redshift
($z_R \ga 8$) the evolution of the gas temperature was also determined by
Compton cooling off the CMB and photo-ionization heating \citep{Hui03}.
In overdense regions, the gas eventually experienced adiabatic heating as
it broke away from the Hubble flow and condensed under the local gravitational 
pull. This effect will modify our results only on the smallest scales. Even 
though filaments and pancakes may form at these high redshifts, the
structures are typically not virialized. One (or two) dimensions may
collapse early but other directions still expand and so the filamentary gas
experiences both adiabatic heating and cooling \citep{Hui97}. Numerical 
simulations of the intergalactic medium indicate that our approximate 
treatment holds for most of the gas mass 
\citep[see e.g., Fig. 1 and top left panel of Fig. 2 in][]{Gardner03}. 

In our simplified treatment, we assume instantaneous reionization of
hydrogen and ignore reionization of helium. Gradual photo-ionization of
hydrogen would keep the IGM at a high temperature for a longer period of
time before cooling processes begin to work. We nevertheless include
recombinations of the ionized IGM and the heat input from subsequent
photo-ionizations.  If \ion{He}{1} reionization occured at a similar
redshift as \ion{H}{1}, it will heat the IGM to a somewhat higher
temperature; however, the temperature of the reionized IGM is uncertain by
an even larger factor due to the unknown spectrum of the ionizing
radiation.  \ion{He}{2} reionization is expected to heat the IGM at a lower
redshift, $z \sim 3$--$4$ (Zheng et al. 2004; Theuns et al. 2002), for
which the characteristic mass scale, $M_{\star}$\footnote{The
characteristic mass scale is implicitly defined through the relationship
$\sigma(M_{\star}) = \delta_c(z_G)$.}, above which the abundance of dark
matter halos is exponentially suppressed, is higher than the mass scale
affected by the high IGM temperature. These circumstances reduce the
influence of \ion{He}{2} reionization on the galaxy power-spectrum at low
redshifts.

\begin{figure}
\plotone{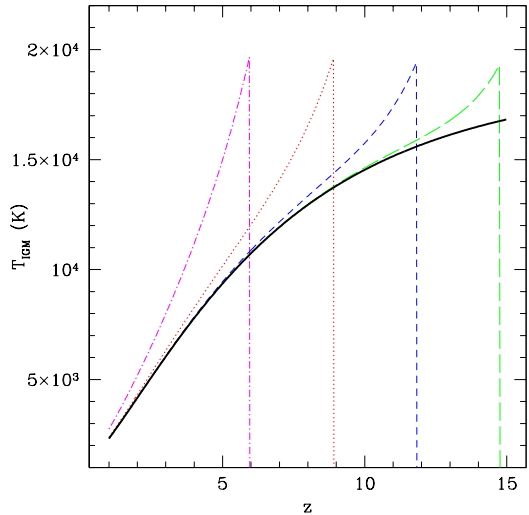}
\caption{The redshift evolution of the IGM temperature for several reionization redshifts: 
$z_R = 6$ (purple, dot-dashed), $z_R = 9$ (red, dotted), $z_R = 12$ (blue, dashed), $z_R = 15$ 
(green, long dashed) and the universal evolutionary track (black, solid).}
\label{temp}	
\end{figure}

Figure \ref{temp} shows the redshift evolution of the IGM temperature for
several reionization redshift: $z_R = 6$ (purple, dot-dashed), $z_R = 9$
(red, dotted), $z_R = 12$ (blue, dashed), $z_R = 15$ (green, long dashed)
and the universal evolutionary track (black, solid). When reionization
occurred early, the IGM temperature quickly approached a common
evolutionary track; this effect is evident in Figure \ref{temp}.  At high
redshifts the timescale for photo-heating and Compton cooling is shorter
than the age of the universe, so the balance between these physical
processes determines the common evolutionary track. If reionization
occurred early enough, the $z=4$ IGM temperature will not be sensitive to
fluctuation around $z_R$.

Since reionization was inhomogeneous, there is a probability distribution
of possible IGM temperatures for any region of scale $R$ in the universe.
The actual IGM temperature of a region regulates the formation of galaxies
in it. First, the gravitational force of the galaxy halo must be able to
overcome the baryon pressure resisting collapse. These considerations
determine the Jeans mass, namely the minimum halo mass into which the IGM
gas may assemble.  Second, the gas that collapses into a galaxy must be
able to cool and dissipate in order to fragment and form stars, and so the
cooling time of the gas must be shorter than the dynamical time of the
halo.  In the absence of molecules, this implies that the virial
temperature of the host galaxy must be greater than $\sim 10^4$ K, so that
atomic line cooling will be efficient.

\begin{figure}
\plotone{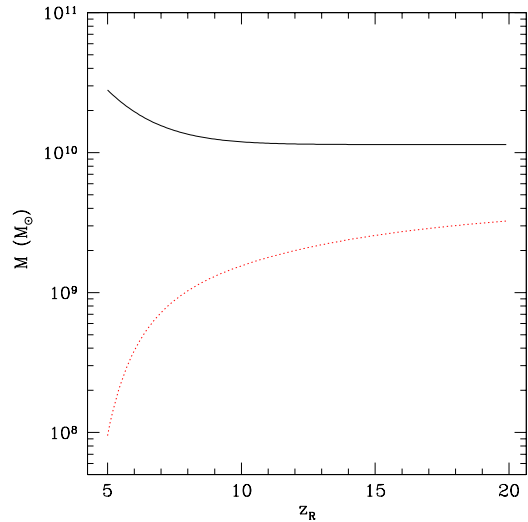}
\caption{Characteristic mass scales: the Jeans mass (black, solid) and the
filtering mass (red, dotted) at $z_G = 4$ for several redshifts of 
reionization, $z_R$.}
\label{masses}	
\end{figure}

The naive Jeans mass overestimates the effects of baryon pressure since the
matter distribution cannot instantaneously adjust to a new thermal and
hydrodynamical state of the gas \citep{Gnedin98}. The adjustment process
occurs on the dynamical timescale of the halo. Incorporating the associated
delay results in a new effective scale, the {\it filtering scale},
which replaces the Jeans scale and can be written in $k$-space as
\begin{equation}\label{filter}
\frac{1}{k^2_F(a)}=\frac{1}{D(a)} \int^a_0 \frac{da'}{a'H(a')} c^2_s(a') 
D(a') \int^a_{a'} \frac{da''}{(a'')^3H(a'')}.
\end{equation}
where $a=1/(1+z)$ is the scale factor and $D(a)$ is the growth factor of
linear density perturbations.  The baryonic power spectrum, $P_b(k)$, which
is strongly influenced by gas dynamics on small scales is then related to
the dark matter power spectrum $P_{dm}(k)$ through
\citep{Gnedin98,Gnedin00b}
\begin{equation}\label{p_b}
   P_b(k,a) = e^{-k^2/k^2_F(a)} P_{dm}(k,a).
\end{equation}
The thermal state of the IGM impacts structure formation through the
suppression of small scale baryonic power and the corresponding increase in
the minimum galaxy mass. Prior to reionization the minimum mass was set by
the cooling time requirement. In reionized regions of the universe the
minimum mass is now set by the filter mass. Figure \ref{masses} shows the
Jeans mass (solid, black) and the filtering mass (red, dotted) at $z_G = 4$
for several values of $z_R$.

The Jeans mass initially decreases as the redshift of reionization
increases.  The wavenumber corresponding to the Jeans length,
\begin{equation}
   k_J(a) = \sqrt{\frac{3}{2}} \frac{a H(a)}{c_s(a)},
\end{equation}
increases for a fixed $a = 1/(1 + z_G)$ as the IGM temperature decreases due to adiabatic
cooling for small $z_R$. For large $z_R$ the IGM temperature approaches a
universal temperature track and Jeans mass becomes a constant because
the IGM temperature for a fixed $z_G$ is independent of $z_R$. The behavior of the
filtering mass is more complicated since the impact of the increased IGM
temperature on the halo mass distribution is delayed. When $z_R$ is close
to $z_G$ the filtering mass can be significantly smaller than the Jeans
mass because of the afforementioned delay. This effect can be seen in
Figure \ref{masses} for small $z_R$. Eventually the altered IGM temperature
is capable of affecting the halo mass distribution and the filtering mass
rises with increasing $z_R$. Since the gas temperature approaches a universal
evolutionary track, the filtering mass curve becomes shallow at high
$z_R$. However, the filtering scale depends on the thermal history and an
earlier reionization implies that the IGM had a high temperature ($\sim
10^4$ K) for a longer time. This causes the filtering mass to monotonically
increase with increasing $z_R$.

Because the temperature of the IGM is inhomogeneous, its effect on galaxy
formation is scale dependent and so it alters the shape of the power
spectrum for the distribution of galaxies. We calculate this effect through
Monte-Carlo simulations of the process. The fluctuations in the galaxy
number density are defined as,
\begin{equation}\label{overdensity}
  \delta_G(R,z_G) = \frac{n(R,z_G) - \bar{n}(z_G)}{\bar{n}(z_G)},
\end{equation}
where $\bar{n}$ is the universal average galaxy number density, ignoring
both fluctuations in the redshift of reionization and modulation of the
galaxy formation process due to the large scale density modes.

Using the Press-Schetcher \citep{Press74} model we calculate the mass
function of galaxies, the cosmic mean of the number density of galaxies
between mass $M$ and $M + dM$ formed at redshift $z_G$, 
\begin{equation}\label{dndM}
  \frac{d\bar{n}}{dM}(M,z_G) =
    \frac{\bar{\rho}}{M}\frac{\delta_c(z_G)}{\sqrt{2\pi}\sigma^3}
    \frac{d\sigma^2}{dM} \exp[-\frac{\delta^2_c(z_G)}{2\sigma^2(M)}].
\end{equation}
Then the mean number density of galaxies in the universe is 
\begin{equation}\label{num}
   \bar{n}(z_G) = \int^{\infty}_{M_{min}} dM \frac{d\bar{n}}{dM}(M,z_G),
\end{equation} 
where $M_{min}$ is the halo minimum mass corresponding to the filter scale,
defined in equation (\ref{filter}).

The ability of a halo to collapse and make a galaxy is altered by the
overdensity of the large-scale region in which it is embbeded.  The
large-scale gravitational evolution of the region modulates the galaxy
formation process by making the linear barrier easier to cross
\citep{Bardeen86}.  Within a region of size $R$ with mass overdensity
$\delta_R$, the modified mass function of halos can be calculated using the
excursion set formalism, also known as extended Press-Schetcher theory 
\citep{Bond91,Sheth99}
\begin{eqnarray}
  \frac{dn}{dM}(M,z_G &;& \delta_R,R) = \frac{\bar{\rho}}{M}\frac{V^L}{V^E}
    \frac{\delta_c(z_G) - \delta_R} {\sqrt{2\pi}
    (\sigma^2(M)-\sigma^2(R))^{3/2}} \nonumber \\ &\times&
    \frac{d\sigma^2}{dM}
    \exp[-\frac{(\delta_c(z_G)-\delta_R)^2}{2(\sigma^2(M) -\sigma^2(R))}],
\label{dndM_R}
\end{eqnarray}  
here $V^L$ is the Lagrangian volume of the region and $V^E$ is its linearly
evolved Eulerian volume. Non-linear evolution of the Eulerian volume will
cause the galaxy number density to dramatically increase on small scales
($R \le 3 \mbox{ Mpc}$). We chose to ignore these non-linear evolutionary
effects in order to clearly display the effect of fluctuations in the
minimum halo mass, which is the focus of this paper.

The number density of galaxies formed in this region is obtained by
integrating equation (\ref{dndM_R}) with the lower mass limit set by
the region's specific mimimum mass,
\begin{equation}
  n(R,z_G) = \int^{\infty}_{M_{min}} dM \frac{dn}{dM}(M,z_G;\delta_R,R).
\end{equation}

The distribution function of the galaxy overdensity, defined in equation
(\ref{overdensity}), can be calculated through the following Monte Carlo 
procedure. 

\begin{enumerate}

\item Randomly generate the local mass overdensity ($\delta_R$) by sampling
the probability distribution function as a Gaussian with zero mean and with
variance given by $\sigma^2(R)$ as defined in equation
(\ref{mass_variance}). If $-1 \le \delta_R \le \delta_c(z_G)$ then proceed,
otherwise resample the PDF.

\item Calculate the fluctuation in the reionization redshift ($\delta z$)
using equation (\ref{delta_z}). If $\delta z < 0$ (the region is a void),
then set $\delta z = 0$, since neighboring overdense regions are expected
to photo-ionize the voids.

\item Assuming the region reionized a redshift $z_R + \delta z$ and
photo-heated to a temperature of $2\times 10^4$ K, calculate the IGM
temperature in the region at the redshift of interest $z_G$, after taking
account of adiabatic cooling due to Hubble expansion, Compton cooling off
the CMB and photo-heating due to the balance between recombinations and
photo-ionizations of the fully-reionized hydrogen. Then calculate the
filtering length and the corresponding minimum galaxy mass, $M_{min}$, in
the region.

\item Use the extended Press-Schetcher formalism in equation (\ref{dndM_R}),
to estimate the number density of galaxies with masses above $M_{min}$ in
the region.  Identify the variance of this galaxy number density divided by
the square of the mean as $1 + \sigma^2_G(R)$.

\item Repeat for other region sizes $R$.

\end{enumerate}

The with statistical moments evaluated according to standard definitions (see
Appendix \ref{moments}). The variance of the distribution can be related to
a galaxy power spectrum,
\begin{equation}\label{s2}
   Var(\delta_G(R))= \sigma^2_G(R) = \int \frac{k^2 dk}{2\pi^2} P_G(k) \tilde{W}^2(kR),
\end{equation}
which can be compared to the standard dark matter power spectrum in order
to quantify the effects of the IGM temperature fluctuations. 

\section{Results}

\begin{figure}
\plotone{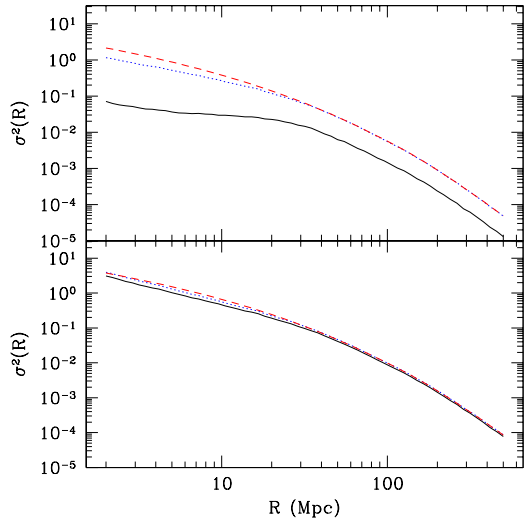}
\caption{The mass variance in the Monte-Carlo simulation of the galaxy
distribution (black, solid curve), Monte-Carlo simulation without including
the fluctuations in $M_{\rm min}$ (blue, dotted curve) and the biased dark
matter mass variance (red, dashed curve) at $z_G = 4$. In the top panel
$z_R = 6$ and in the bottom panel $z_R = 15$.}
\label{variance}	
\end{figure}

Based on the formalism described in the previous section we calculated the
variance and skewness of the galaxy number overdensity. Figure
\ref{variance} shows the mass variance in equation (\ref{s2}) of galaxies
formed at $z_G = 4$ for reionization redshifts of $z_R = 6$ (top panel) and
for $z_R = 15$ (bottom panel). The solid black curves are the Monte-Carlo
simulated mass variances including the effects of inhomogeneous
reionization, while the red dashed curves are the linear dark matter mass
variances produced by CMBFAST\footnote{http://www.cmbfast.org/} and
adjusted for linear bias. The bias parameter is averaged over the mean mass
function in equation (\ref{dndM}),
\begin{equation}\label{bias}
   \bar{b} = \frac{1}{\bar{n}(z_G)} \int^{\infty}_{M_{min}} dM b(M)
   \frac{d\bar{n}}{dM}(M,z_G),
\end{equation}
where the Eulerian linear bias is given by \citep{Mo96}
\begin{equation}\label{mo_white}
   b(M) = 1 + \frac{\delta^2_c(z_G)/\sigma^2(M) - 1}{\delta_c(z_G)};
\end{equation}
for large mass halos this formula is accurate \citep{Sheth01}.
The linear bias is calculated by expanding the extended Press-Schetcher
mass function, equation (\ref{dndM_R}), to first order in $\delta_R$.
The mass averaged bias, equation (\ref{bias}), can be altered to account
for the fluctuations in the minimum halo mass due to the local mass
overdensity. The minimum halo mass can be linearized as
\begin{equation}
  M_{min} = \bar{M}_{min} + M' \delta_R,
\end{equation} 
where $M'\equiv (dM_{min}/d\delta_R)$, $\bar{M}_{min}$ is the average
minimum halo mass, then the mass averaged linear bias will have the extra
term
\begin{eqnarray}
    \frac{1}{\bar{n}(z_G)}& &\int^{\bar{M}_{min}}_{\bar{M}_{min} + M'
    \delta_R} dM b(M) \frac{d\bar{n}}{dM}(M,z_G) \nonumber \\ &\approx& -
    b(\bar{M}_{min}) \frac{1}{\bar{n}(z_G)}
    \frac{d\bar{n}}{dM}(\bar{M}_{min},z_G) M' \delta_R.
\end{eqnarray}

For $z_R = 6$ the fluctuations in the minimum mass will be anti-correlated
with the standard Lagrangian biasing effect since the minimum halo mass
curve has a positive slope (see Fig. \ref{masses}). Physically this means
that an overdense region which reionized earlier than average, will have a
larger minimum halo mass because the halo mass distribution has had a
longer amount of time to reflect the photo-heated IGM gas pressure. If $z_R
= 15$ the effects of the fluctuation in the minimum halo mass will be
weakly anti-correlated with the Eulerian linear bias. In the case of early
reionization the IGM temperature approaches a universal evolutionary
track. This makes the minimum halo mass curve shallow and reduces the
effect of inhomogeneous reionization.  This feature is visible in Figure
\ref{variance}; the $z_R = 6$ mass variance is more suppressed than the
$z_R = 15$ mass variance on large scales where the linear bias is most
applicable. The $z_R = 15$ mass variance is suppressed by approximately $10
\%$ on large scales, while the $z_R = 6$ mass variance is suppressed by
just less than a factor of 3.  Additionally, the $z_R = 6$ mass variance
drastically changes its shape near $R \sim 20$ Mpc. On this scale the
fluctuations in the redshift of reionization begin to be substantial (see
Fig. \ref{dz}). These effects would become more prominent, especially in
the $z_R = 6$ case, if we decided to consider $z_G = 5$, since the slope of
the $M_{\rm min} - z_R$ curve would become steeper. Likewise the effects
would be more modest if we chose $z_G = 3$. In Figure \ref{variance} we
include the output of the Monte-Carlo code ignoring the effect of the scale
dependent fluctuations in the minimum halo mass (blue, dotted curve). This
comparison demonstrates that differences between the Monte-Carlo code
curves and the biased dark matter mass variance are not dominated by
numerical effects.

The effects of radiative feedback on galaxy formation were assumed to be
restricted to the small scales on which individual galaxies could impact
the evolution of their neighbors. We have found that the radiative feedback 
produced by the process of reionization changes even the large scale 
($> 100$ Mpc) amplitude of the galaxy mass variance (see Fig. \ref{variance}).
The ionizing sources are highly clustered \citep{Furlanetto04b,Wyithe05c}, so the
\ion{H}{2} regions during reionization are much larger than the volume that
an individual galaxy could ionize. It is the large bias of the ionizing
sources which allows the radiative feedback effects to influence the
large scale galaxy power spectrum.

The proposed Cosmic Inflation Probe ({\it CIP}) satellite \citep{Melnick05}
aims to measure the galaxy power spectrum down to $k \sim 1 \mbox{
Mpc}^{-1}$ in order to constrain inflationary parameters. Currently the
best constraints on $n$ and $\alpha$ come from a joint analysis of WMAP
data and SDSS Ly$\alpha$ forest data \citep{Seljak05}; the best single
parameter $68.32\%$ confidence level constraints are $\Delta n = \pm 0.02$
and $\Delta \alpha = \pm 0.01$. Once the covariance between $n$ and
$\alpha$ is included, the constraints considerably worsen \citep[see Figure
3 in][]{Seljak05}.  The prospective constraints from {\it CIP} are much
finer, $\Delta n = \pm 0.002$ \citep{Melnick05}.

\begin{figure}
\plotone{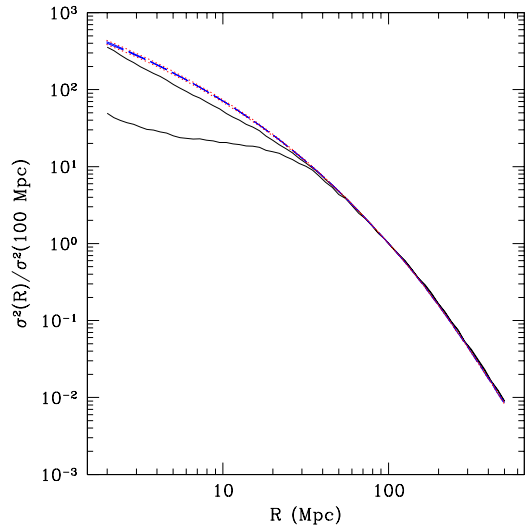}
\caption{The normalized $\sigma^2(R)$ for $z_R = 15$ and $z_R = 6$ (upper
and lower solid black curves, respectively); the limiting cases of the best
current constraints of $n$ (red, dotted curves) and $\alpha$ (blue, dashed
curves).}
\label{params}	
\end{figure}

In Figure \ref{params} we show $\sigma^2(R)$, normalized to $1$ at $R = 100
\mbox{ Mpc}$, for $z_R = 15$ and $z_R = 6$ (solid, black curves) and the
two limiting cases of the best constraints from \cite{Seljak05}; $n = 1 \pm
0.02$, $\alpha = 0$ (dotted, red curves) and $n = 1$, $\alpha = 0 \pm 0.01$
(dashed, blue curves). The effects of inhomogeneous reionization clearly
dwarf the slight changes caused by varying the inflationary
parameters. Therefore any program which seeks to constrain cosmological
information from high redshift galaxy surveys must include the impact of
inhomogeneous reionization on the shape of the galaxy power spectrum.

The imprint of reionization on the high redshift galaxy power spectrum is
strong because we assumed that galaxies which reside in $M_{\rm min}$ halos
will be observed.  If the survey flux limit corresponded to a much larger
halo mass, the imprint of inhomogeneous reionization on the galaxy power
spectrum would have been reduced\footnote{The target minimum halo mass for
CIP is $3 \times 10^{11} M_{\sun}$ \citep{Komatsu05}.}.  However, galaxies
which are only a few times more massive than $M_{\rm min}$ are still
assembled through mergers of lower mass galaxies and their luminosity
should be significantly affected by a depletion in the star formation rate
of their building blocks of mass $M_{\rm min}$. Additionally, supernova
feedback may deplete the progenitors of their baryonic gas, and then the
inhibited gas accretion may significantly prevent any subsequent star
formation in these halos (e.g. Dekel \& Silk 1986). Since the effect of reionization could be
larger by orders of magnitude than the allowed range of variations due to
uncertain inflationary parameters, it may not be practical to confine a
search to sufficiently massive halos where the inflationary variations
dominate. The exact effect that photo-heating of the IGM will have on halos
with masses above $M_{\rm min}$ is difficult to estimate analytically.  The
analysis which leads to the filtering mass or similarly the Jeans mass, is
based on linear theory. Although CIP's minimum halo mass is significantly
above the minimum halo mass predicted by the filtering scale, the
inhomogeneous effects of reionization are substantially larger (see
Fig. \ref{params}) than the effect of the tilt and running of the
primordial power spectrum that CIP is designed to measure. It is possible
that inhomogeneous reionization may bias CIP's measurement of the linear
galaxy power spectrum and the inflationary parameters.  A comprehensive
calculation of our effect requires a full hydrodynamical simulation of
galaxy formation in a box large enough to account for the biasing by large
scale density perturbations which cause inhomogeneous reionization
\citep[see][for an overview of the effects of box size on simulations of
reionization]{Barkana04}. The inclusion of a homogeneous UV ionizing
background in simulations or semi-analytic models of galaxy formation will
not lead to the proper bias nor will it include the crucial effect of
inhomogeneous reionization which alters the shape of the galaxy power
spectrum.

The distortion of the galaxy power spectrum may be used as a new probe of
inhomogeneous reionization.  Even if a wide field survey like {\it CIP}
might not have the sensitivity to observe galaxies in halos as small as
$M_{\rm min}$, the {\it James Webb Space Telescope}
(JWST)\footnote{http://www.jwst.nasa.gov/} with $\sim 1$ nJ sensitivity
most certainly will. A deep exposure (similar in concept but deeper than
the Hubble Deep Field) of a field with dimensions of $5\times 5 \mbox{
arcmin}^2$ will observe $\sim 10^6$ galaxies between $z = 3$ and $z =
5.5$. The large density of galaxies (assuming source confusion is not a
significant problem when galaxies are sliced in redshift) should eliminate
the Poisson sampling errors which typically prevent high redshift galaxy
counts from placing strong constraints on cosmological parameters.

\section{Discussion}

The process of reionization is intrinsically inhomogeneous due to the
modulation introduced by large scale structure on the formation of ionizing
sources. Regions which are overdense form structure earlier than average
and correspondingly reionize earlier than average.  As the regions are
photo-ionized they get photo-heated to a characteristic temperature of
$\sim 2 \times 10^4$ K, and subsequently cool.  Regions that reionized
earlier than average experienced suppressed galaxy formation for a longer
amount of time. Consequently, the minimum mass of dark matter halos that
can accomdate star formation, $M_{min}$, is higher in such regions.  This
feedback of reionization on the minimum halo mass alters the galaxy power
spectrum.  Figure \ref{params} summarizes the effects of inhomogeneous
reionization on the distribution of galaxies at high redshifts.

The filtering scale differs from the Jeans scale since it involves an
average over the thermal history. Immediately following the photo-heating
of the IGM, the filtering scale lags behind the Jeans scale. The delay is
displayed in Figure \ref{masses}, where the filtering mass rises initially
rapidly and later more gradually as the redshift of reionization increases,
while the Jeans mass declines at first and then approaches a constant. If
reionization occurred at $z_R = 6$, the overdense regions which reionize
earlier would have a larger $M_{min}$ than average, reducing the number
density of galaxies. This is anti-correlated with the well-known biasing
effect of large scale density modes which produces more galaxies in
overdense regions. If reionization occured at $z_R = 15$, the marginal
increase in the minimum mass for overdense regions is more modest since the
photo-heated IGM quickly approached a universal evolutionary track

The new effect we calculated is scale dependent, since the fluctuations in
the redshift of reionization are larger on small scales (see
Fig. \ref{dz}). This consequently implies that the proper interpretation of
the high redshift galaxy power spectrum will require a detailed
understanding of reionization. Without adequate knowledge of the IGM
temperature history, it may not be possible to constrain the inflationary
parameters (through the spectral index and the running of the spectral
index of scalar perturbations) to a high precision. Conversely, the shape
of the galaxy power spectrum at high redshifts contains crucial information
about the scale dependence of the thermal history of the IGM during and
after the epoch of reionization.  Previous attempts to measure the IGM
temperature through its imprint on the small scale power spectrum of the
Ly$\alpha$ forest \citep[see][and references within]{Zaldarriaga02} are not
easy to implement \citep{Lai05}. Since the process of galaxy formation is
inherently non-linear, the IGM temperature flucuations are amplified and
become easier to detect.  A deep exposure by {\it JWST} or a wide field
galaxy survey like {\it CIP} may prove to be effective new probes of the
epoch of reionization.

\acknowledgements 

We thank Adam Lidz for useful conversations and Gary Melnick for providing
information on the Cosmic Inflation Probe. This work was supported in part
by NASA grants NAG 5-13292, NNG05GH54G, and NSF grants AST-0071019,
AST-0204514 (for A.L.).

\begin{appendix}

\section{Moments Estimators} \label{moments}

In this work we Monte-Carlo several different physical quantities 
(mass overdensity, reionization redshift fluctuation, filtering length and 
galaxy overdensity) in order to determine the quantities' probability 
distribution function (PDF). The PDFs can be completed described by 
the infinite set of their statistical moments. In this work we will only
measure the lowest three moments.
The first moment, the mean, is defined as
\begin{equation}
  Mean({x_i}) = \bar{x} = \frac{1}{N} \sum_i x_i,
\end{equation}
the variance is defined as
\begin{equation}
  Var({x_i}) = \sigma^2 = \frac{1}{N-1} \sum_i (x_i-\bar{x})^2.
\end{equation}

\end{appendix}

\end{document}